\documentstyle[aps,preprint,prc]{revtex}

\begin{document}

\begin{center}
{\Large
Scaling Law in Cluster Decay}\\
\vspace{.9cm}
Mihai Horoi$^{1,3)}$,
B. Alex Brown$^{1)}$ and Aurel
Sandulescu$^{2,3)}$ \\

\vspace{.9cm}

{\it $^{1)}$National Superconducting Cyclotron Laboratory and\\
Department of Physics and Astronomy,\\
Michigan State University, East Lansing, MI 48824 USA\\
\vspace*{0.2cm}
$^{2)}$Institut f\"ur Theoretische Physik, J. W. Goethe
Universit\"at,\\
 W-6000 Frankfurt/Main, Germany\\
\vspace*{0.2cm}
$^{3)}$Institute of Atomic Physics, Bucharest, Romania\\}
\end{center}

\vspace{.3cm}

\begin{abstract}
A recently proposed scaling law for the decay time of alpha
particles is generalized for cluster decay. It is shown that for
the decay of even-even parents, $logT_{1/2}$ depends linearly
on the
scaling variable S=(Z$_{c}$Z$_{d})^{0.6}/\sqrt{Q_{c}}$ and on the
square root of the reduced mass of cluster and daughter.
\vspace*{.7cm}

\end{abstract}

\newpage

In the last years alpha decay systematics
have been theoretically reinvestigated
\cite{BAB}-\cite{HNH}
partly as a result
of the special interest in the studies of
superheavy nuclei and cluster decay. In the same time the cluster
decay data has been accumulated \cite{Pr} with a sufficiently
high precision so that their systematics can be now studied
almost {\it model independently} with
guidance to the alpha decay. All the predictions or analysis of the
cluster decay data given up to now \cite{SS}-\cite{BM} are based on
models. The prescriptions of these models and/or the physical meaning of
the parameters used by them have been subjected to criticism. In
this letter we present a {\it model independent} description of the
accumulated data pointing out the most important variables which scale
the cluster decay probabilities.

The earliest phenomenological connection between the half lives
and the $Q$-values of the alpha decays of radioactive series was
proposed by Geiger and Nuttall \cite{GN}

\begin{equation}
logT_{1/2} = a Q_{\alpha}^{-1/2} + b\ ,
\end{equation}

\noindent
which proved to be very useful for the prediction of alpha decay
half lives. The rule of Geiger and Nuttall has the disadvantage
that the $a$ and $b$ parameters are dependent on the isotope chain.
A universal
scaling law for alpha decay half lives of the even-even parents
has recently been proposed
\cite{BAB}

\begin{equation}
logT_{1/2} = 9.54 \frac{Z_{d}^{0.6}}{\sqrt{Q_{\alpha}}} - 51.37\ ,
\end{equation}

\noindent
where Z$_d$ is the charge of the daughter. It was shown that the
known \cite{BMP1} alpha decay $logT_{1/2}$ of the even-even nuclei
with Z$\ge$76 stay on this universal line with a rms deviation
of 0.33.

It is of interest to know if similar scaling law(s) exist(s) for
cluster decay:

\begin{equation}
^{A_p}Z_p \rightarrow ^{A_d}Z_d + ^{A_c}Z_c
\end{equation}

\noindent
where the subscripts $p$, $d$, $c$ refer to the parent,
daughter and cluster, respectively.
We look
to even-even parents and clusters
with the hope that the structure effects
\cite{GHSGS} are limited and only the collective dynamics
dominates the process. Relation (2) has been
compared to what one obtains from that obtained from
classical
Gamow formula for the alpha decay constant

\begin{equation}
\lambda = \omega P_{\circ} P\ ,
\end{equation}

\noindent
where $\omega$ is the frequency with which the alpha particle
exists at the barrier, $P_{\circ}$ is the preformation probability
(assumed to be constant
in Ref. \cite{BAB}) and $P$ is the barrier penetration factor
assuming a square well plus a Coulomb potential for
the radial dynamics. One obtains

\begin{equation}
logT_{1/2} = C_{\circ}+
2\cdot log2\cdot \frac{Z_{d} Z_{c}}{\sqrt{Q_{\alpha}}} e^{2} \sqrt{2
\mu /\hbar^{2}} [arccos(x) - x \sqrt{1-x^{2}}]\ ,
\end{equation}

\noindent
where
\begin{equation}
x = \sqrt{\frac{Q_{c}(
R_{c} + R_{d})}{Z_{c} Z_{d} e^{2}}}\
\end{equation}

\noindent
and where $Q_c$ is the cluster decay Q-value and
the $R$ are the equivalent hard-sphere charge radii
and $C_{\circ} = log(ln2/\omega P_{\circ})$.

The scaling law (2) is not obvious from Eq. (5), but for $x \leq 0.8$,
$logP$ behaves approximately
linearly as a function of
Z$^{0.6}_{d}/\sqrt{Q_{\alpha}}$. This is in fact the region of
interest for the alpha decay of heavy nuclei.
The above
analysis indicates that the scaling variable is
(Z$_{\alpha}$Z$_{d})^{0.6}/\sqrt{Q_{\alpha}}$. As a consequence, we
have looked for the behavior of the known experimental data on cluster
decay (see e.g Table 1 from Ref. \cite{SG}) as a function of the
cluster scaling variable

\begin{equation}
S = \frac{(Z_c Z_d)^{0.6}}{\sqrt{Q_c}}\ .
\end{equation}

\noindent
The data are
presented in Fig. 1. There are only 3 known "chains" of cluster
decay with more than one element: $^{14}$C ($logT_{1/2}^{exp}$=11.02 from
$^{222}$Ra \cite{P85}, $logT_{1/2}^{exp}$=15.9 from $^{224}$Ra \cite{P85} and
$logT_{1/2}^{exp}$=21.33 from $^{226}$Ra \cite{BA86}), $^{24}$Ne
($logT_{1/2}^{exp}$=20.41 from $^{232}$U \cite{BPL}, $logT_{1/2}^{exp}$=24.64
from $^{230}$Th \cite{TR85} and $logT_{1/2}^{exp}$=25.24 from $^{234}$U
\cite{MO89}) and $^{28}$Mg ($logT_{1/2}^{exp}$=21.68
from $^{236}$Pu \cite{OG90}
and $logT_{1/2}^{exp}$=25.75 from $^{234}$U \cite{MO89}).
The
experimental data concerning $^{14}$C and $^{24}$Ne cluster decay
clearly shows a linear dependence as function of the scaling
variable S.
This analysis indicates a scaling law
for the
cluster decay (alpha included) similar to Eq. (2)

\begin{equation}
logT_{1/2} =C_1 (S-7) + C_2\ .
\label{eq:scal1}
\end{equation}

\noindent
The constant "seven" is subtracted from S in this equation simply
so that the parameter $C_2$ has a numerical value which is
close the actual experimental values shown in Fig. 1.
The coefficients C$_1$ and $C_2$ can be extracted from the fit of
the experimental data. The values of the C$_{1}$ parameters
are 6.3 for $^{4}$He,
17.4 for $^{14}$C, 20.7 for $^{24}$Ne and 27.1 for $^{28}$Mg. The corresponding
values of C$_{2}$ are: -7.3, 8.0, 19.1 and 21.7.

It is interesting to examine if there are some correlations between
the C$_1$ and C$_2$ coefficients corresponding to different cluster
decays. Eq. (5) suggests an additional dependence on the
masses. In the alpha decay case the dependence on $\sqrt{\mu}$
(the reduced mass) is very small: 0.5 \% for the mass of the
daughter (A$_d$) in the range 150 to 250.
For clusters
heavier than alpha the dependence on $\sqrt{\mu}$ is very important.
The analysis of the heavy
cluster decay case is complicated by the fact that the preformation
probability (prescission probability \cite{SS}) plays a more important role
in the majority of the theoretical models \cite{GHSGS}-\cite{BBB}.
There are also models \cite{Pr,BM}
which assure the prescission probability to be unity.

Guided by Eq. (5)
and by the fact that some models \cite{PGDIMS,SS} indicate a
$\sqrt{\mu}$ dependence of the prescission part also, we have
plotted in Fig. 2 the C$_1$ and C$_2$ coefficients as a function of

\begin{equation}
\sqrt{\mu} = \sqrt{\frac{A_{c} \cdot  A_{d}}{A_{c} + A_{d}}}\ .
\end{equation}

\noindent
For the plotting purpose only we have used A$_{d} = 208$
neglecting the small variation due to the different
daughter masses.
One can clearly see a linear dependence on
$\sqrt{\mu}$ of these coefficients
(the fitted lines are given by $C_{1} = 6.3 \sqrt{\mu} - 6.2,\
C_{2} = 9.8 \sqrt{\mu} - 26.9$).

An alternative
fit of the cluster decay data,
in the spirit of those models \cite{Pr,BM}
which consider this preformation probability equal to unity,
can
be performed with the help of Eq. (5). We have taken
$R_{c}=0.0354 A_{c} + 2.008\ (fm)$
for the cluster radius
which empirically reproduces the charge radii of the light clusters,
and $R_{d} = r_{\circ}A_d^{1/3}$ for the daughter radius.
The experimental data  were fitted by using the two parameters
$C_{\circ}$ and $r_{\circ}$.  A 0.64 rms deviation from
the experimental values has
been obtained and the following values for the fitted parameters:
$C_{\circ} = -23.1$, $r_{\circ} = 0.976$ fm. The $r_{\circ}$
value is 0.25 fm smaller than the typical values for heavy
nuclei, and this make the touching radius 1 - 1.5 fm smaller.
This is an unreasonable reduction, but
the "extra-penetrability" could simulate the  preformation
probability  (the -23.1 value is consistent with a preformation
probability $P_{\circ}$ of about unity contributing to $C_{\circ}$).

Assuming that the preformation probability is different from unity,
our analysis
indicates that not only the postscission but the
prescission dynamics
also is dominated by the square root of the
reduced mass. It is
interesting to compare this conclusion with the prescriptions
presented by different models. The present analysis is in accord
with the prescription of Ref. \cite{PGDIMS} and \cite{SS}. The
work of Blendowske and Walliser \cite{BW} indicates a linear
dependence on A$_c$ (the cluster mass) of the prescission
probability (the spectroscopic factor in Ref. \cite{BW}), differing
from the present conclusion. Barranco, Broglia and Bertsch
\cite{BBB} have obtained in their superfluid tunneling model
a dependence of the prescission probability on the number of steps
to the scission (an extra dependence of the gap parameter entering their
formula on the mass of the
parent nucleus does not affect this analysis).
 This number is very close to the reduced mass
$\mu$
 and not to $\sqrt{\mu}$, again differing from the our
findings.

The above analysis indicates a  {\it model independent} law for
the whole body of cluster decay data of the following form:

\begin{equation}
logT_{1/2} = (a_{1} \mu^{x} + b_{1})\left[ (Z_{c} Z_{d})^{y}/\sqrt{Q}-7
\right]
+ (a_{2} \mu^{x} + b_{2})\ .
\end{equation}

\noindent
A fit of the 119 alpha decays\cite{BAB} and 11 cluster decays\cite{SG}
from even-even parents has been done. Besides the 8 "in chain" cluster data
considered in Fig. 1, 3 "single" cluster data have been taken into the fit:
 $^{20}$O from $^{228}$Th
($logT_{1/2}^{exp}$=20.9 \cite{BNP}), $^{32}$Si from
$^{238}$Pu ($logT_{1/2}^{exp}$=25.3 \cite{WA89}) and
$^{34}$Si from $^{242}$Cm ($logT_{1/2}^{exp}$=23.2 \cite{pcT}).
The fit result
 gives $a_{1} = 9.1$, $b_{1} = -10.2$, $a_{2} = 7.39$, $b_{2} =
-23.2$, $x = 0.416$ and $y = 0.613$ with a 0.34 rms deviation of $logT_{1/2}$.
Considering the important parameter $x$, a range of values from 0.4 to
0.6 can be obtained depending upon the various subsets of data used in
the fit.
A 0.58 rms is extracted for the heavy clusters only, which represents a fairly
good description of the data if one has in mind that the largest deviation
comes from $^{34}$Si ($logT_{1/2}$=24.45 as compared with the 23.2 experimental
value).
This may indicate that the extrapolation of Eq. (10)
to heavier clusters must be taken with caution.
One would also like to see  independent confirmation of the $^{34}$Si
experimental investigations\cite{pcT}.
The apparent breakdown of this scaling law when going from cluster decay to
fission could be understood by the fact that in the latter case the
dynamics is not dominated by the Coulomb potential but
by the collective potential up to the scission point. We compared our
formula with model dependent results for heavier cluster decay like
$^{48}$Ca from $^{256}$No. Our result is $logT_{1/2} = 27.9$ while the
result from Ref. \cite{PSGMG} is significantly smaller
($logT_{1/2} = 18.9$). Experimental information in this mass range are
crucial.

The scaling law, Eq.(10),
can be straightforwardly used to produce tables with cluster decay
half live predictions similar with those in Ref. \cite{PSGMG}.
Input parameters are the mass and charge
numbers of cluster and daughter and the Q-value of the reaction.
A detailed search through all the possible decays of the parents
with $82 \leq Z_{p} \leq 106$ and clusters with
$2 \leq Z_{c} \leq 20$ shows the possibility to obtain
experimental data for the
decay of new clusters in this region: e.g  $^{12}$C from $^{220}$Ra
($logT_{1/2} = 10.4$) and from
$^{222}$Th ($logT_{1/2} = 10.08$), $^{18}$O from $^{226}$Th
($logT_{1/2} = 17.75$), etc.
To
select these cases we have used similar constraints as in Ref.
\cite{PSGMG}, namely $logT_{1/2} \leq 28$ and $logT_{1/2} -
logT_{1/2}(\alpha) \leq 18$. Only those nuclides for which the
experimental masses are known \cite{WA} have been used.

One can try to test the $\sqrt{\mu}$ behavior in
decays for which the daughter is different from the $^{208}$Pb
region.  The neutron deficient A $\approx$ 120 region is particularly
interesting. For
example,
the decay of $^{118}$Ba into a $^{12}$C cluster  and
$^{106}$Sn. Eqs. (10) together with
an experimentally extracted Q-value,  $Q_c$=15.10 MeV
\cite{PSGMG},
gives
$logT_{1/2}$ = 18.0.
This extrapolation of the scaling law (8) to light parents gives significantly
lower half lives as compared with other {\it model dependent} treatments
(e.g. $logT_{1/2}$=21.3 in Ref.
\cite{PSGMG}). Further experimental tests are required to validate one
of these approaches.

Eq. (10)
represents the first {\it model independent} description of all known
cluster decay data. The parameters $a_{1},\ b_{1},\ a_{2},\
b_{2}$, $x$ and $y$ contain information on the dynamics of the decay.
The actual theoretical models describing the cluster decay data are
rather crude. Often their  parameters  loose
their physical meaning, as for the unphysically small $r_{\circ}$
discussed above or
e.g. the use of a zero-point motion energy even in the asymptotic
region \cite{PSGMG}.
In our approach we have emphasized  the most important variables
(S, $\sqrt{\mu}$) scaling the experimental data. We expect  this new
approach to be an important step toward a theoretical
description of the cluster
decay.

In conclusion, we have obtained a new scaling law for the alpha and
 cluster decay
of the even-even heavy nuclei.
The scope of these
scaling laws is to describe the regularities of the data, to put
in evidence the peculiar behavior with respect to these
regularities, to reveal the most important parameters entering
the theoretical models and to guide new prediction for the cluster decay.
New experimental data are necessary to further support the
present analysis.

\vspace{1.2cm}
\noindent

MH and
BAB would like to acknowledge
support from the Alexander von Humboldt Foundation and NSF grant 90-17077.


\newpage

\begin{center}
{\bf Figure captions}
\end{center}

\vspace{1cm}

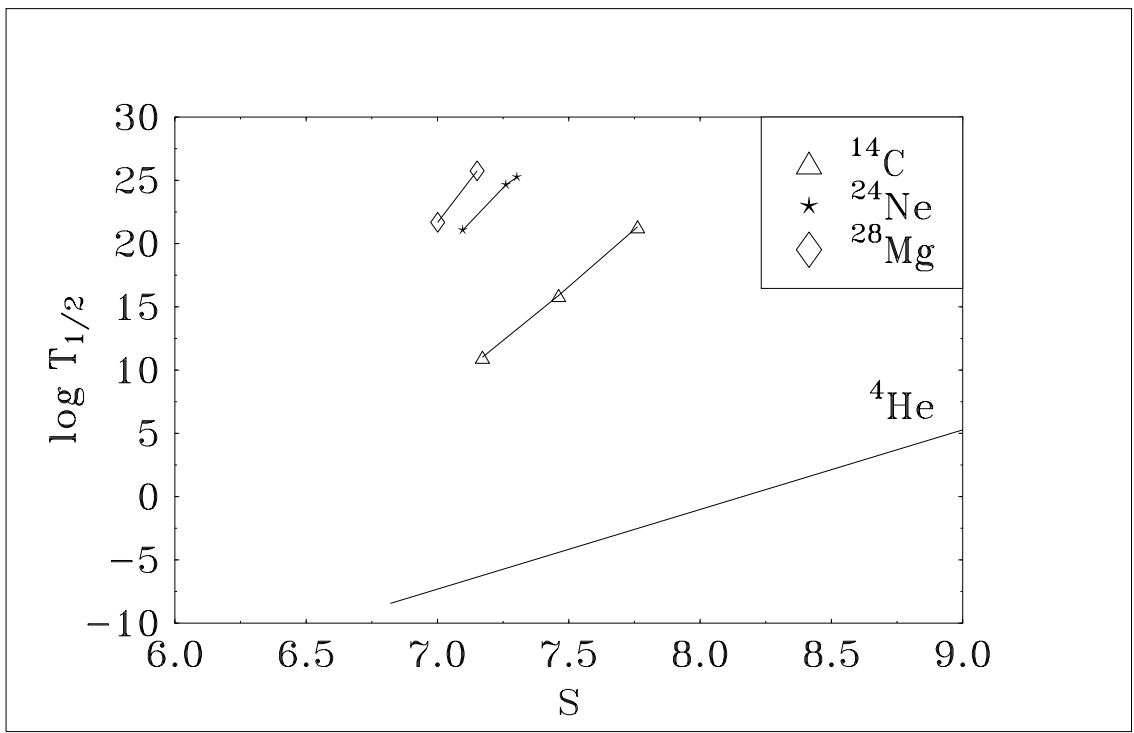
{\bf Figure 1} \  Experimental data for $logT_{1/2}$ (sec) of the
cluster decay of even-even parents as functions of the scaling
variable S (Eq. (7)). The line noted by $^{4}$He is given by Eq. (2)
which represents the best linear fit to the experimental data.
Other lines are drawn to guide the eye.
\\

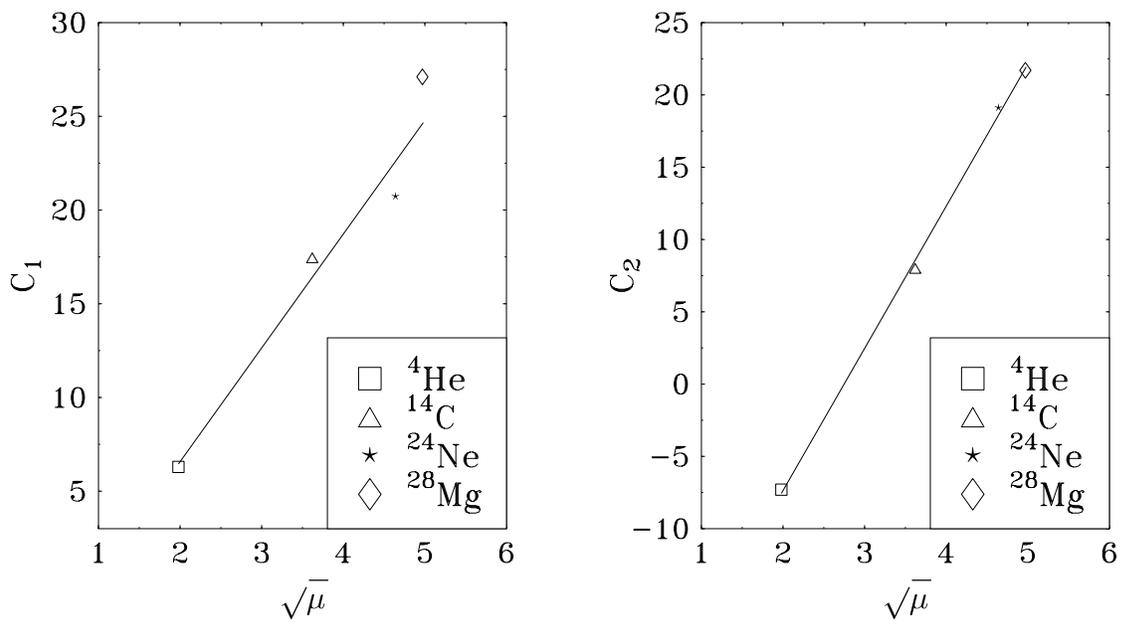
{\bf Figure 2} \ C$_1$ and C$_2$ coefficients entering Eq. (\ref{eq:scal1})
as function of $\sqrt{\mu}$ defined in Eq. (9). Lines represent the best
linear fit.

\end{document}